\documentclass[useAMS,usenatbib]{mn2e}
\usepackage{graphicx}
\usepackage{amssymb}
\usepackage{amsmath}

\title[IMBH Formation in the Inflationary Cosmology]
{Formation of intermediate-mass black holes as primordial black holes in
the inflationary cosmology with running spectral index}

\author[Kawaguchi et al.]
{Toshihiro Kawaguchi$^{1}$,\vspace{0.35 cm} Masahiro Kawasaki$^{2,3}$, Tsutomu
Takayama$^{2,\star}$,\\ 
\hspace{-0.1 cm}{\LARGE{\rm Masahide Yamaguchi}$^{1}$}
{\LARGE{\rm and Jun'ichi Yokoyama}$^{4,3}$}\vspace{0.25 cm}\\
$^{1}$Department of Physics and Mathematics, Aoyama Gakuin University,
Sagamihara 229-8558, Japan\\
$^{2}$Institute for Cosmic Ray Research (ICRR), The University of Tokyo,
Kashiwa 277-8582, Japan\\
$^{3}$Institute for the Physics and Mathematics of the Universe,
The University of Tokyo, Kashiwa 277-8582, JAPAN\\
$^{4}$Research Center for the Early Universe (RESCEU), Graduate School
of Science, The University of Tokyo, Tokyo 113-0033, Japan }

\begin{document}

\pagerange{\pageref{firstpage}--\pageref{lastpage}} \pubyear{2007}

\maketitle

\label{firstpage}

\begin{abstract}
Formation of primordial black holes (PBHs) on astrophysical mass scales
is a natural consequence of inflationary cosmology if the primordial
perturbation spectrum has a large and negative running of the spectral
index as observationally suggested 
today, because double inflation is
required to explain it and fluctuations on some astrophysical scales are
enhanced in the field oscillation regime in between.  It is argued that
PBHs thus produced can serve as intermediate-mass black holes (IMBHs)
which act as the observed ultraluminous X-ray sources (ULXs) by choosing
appropriate values of the model parameters in their natural ranges.  Our
scenario can be observationally tested in near future because the mass
of PBHs is uniquely determined once we specify the values of the
amplitude of the curvature perturbation, 
spectral index and its running on large scales.
\end{abstract}


\section{Introduction}
Ultraluminous X-ray sources (ULXs; Makishima et al. 2000) 
are characterized by their luminosity greater
than $\sim10^{39}{\mathrm{erg~s}}^{-1}$ (well above the Eddington 
luminosity of a neutron star) 
and off-nuclear location in 
 nearby galaxies (Fabbiano 1989; Colbert \& Mushotzky 1999). 
In addition, 
X-ray variability on various timescales 
are found for 
some ULXs (Matsumoto \& Tsuru 1999; Ptak \& Griffiths 1999; 
Kaaret et al. 2001; Kubota et al. 2001). 
Luminosity, time variability and 
X-ray spectra of ULXs 
indicate that ULXs are accreting black holes, rather than young 
supernova remnants \citep{Kaaret:2000sh}.
Although association of ULXs with active star-forming
regions are clearly shown (Matsushita et al. 2000; Zezas et al. 2002), 
the physical reason behind the association,
and moreover the origin of ULXs, are still unclear. 

The number density of ULXs can be roughly estimated as follows.
Late-type/starburst galaxies hosting numerous compact
X-ray sources (mostly high-mass X-ray binaries) within about
$30{\mathrm{Mpc}}$ from the Milky Way are listed in
\citep{Grimm:2002ta}.
The number of these X-ray sources with each luminosity
greater than $2\times10^{38} {\mathrm{erg~s}}^{-1}$ is about 100. 
Grimm et al.~ also show that the
luminosity function $N(L)$ of compact X-ray sources, which is defined as
a number of X-ray sources with luminosity greater than $L$, is universal
for galaxies if it is normalized proportionally to the star-forming rate (SFR).  
The normalized luminosity function tells that 
among $L>10^{38.5}{\mathrm{erg~s}}^{-1}$ sources 
about one-third have 
$L > 10^{39}{\mathrm{erg~s}}^{-1}$. 
Hence, the number of ULXs with 
$L>10^{39}{\mathrm{erg~s}}^{-1}$ 
will be $\sim$30 within 30\,Mpc. 
Thus, the number density of ULXs is estimated to be
$10^{-3.5} {\mathrm{Mpc}}^{-3}$.

In spite of observational evidences, theoretical explanations of ULXs are
still in dispute.  The most challenging fact is the high luminosity
greater than the Eddington
luminosity of a stellar-mass black hole with mass $\sim10M_\odot$.  For
example, it is argued that sources of ULXs can be standard stellar-mass
black holes with jets or relativistic beaming \citep{Koerding:2001am}.
Another possibility is stellar-mass black holes 
radiating at super-Eddington luminosities due to 
efficient photon leakage from accretion disks 
(e.g., Begelman 2002; Meyer 2004) and/or due to 
super-Eddington accretion rates (``slim accretion disks''; 
Abramowicz et al. 1988).
Detailed computations of slim accretion disks (Kawaguchi 2003) 
indeed account for observed X-ray spectra of ULXs 
(Okajima, Ebisawa \& Kawaguchi 2006; Vierdayanti et al. 2006; 
Ebisawa \& Kawaguchi 2006; Foschini et al. 2006). 

ULXs can also be explained as sub-Eddington accretors 
by assuming black holes heavier than stellar-mass black holes (see 
Miller \& Colbert 2004 and reference therein).  
Moreover, ULXs exhibit quasi-periodic oscillations (QPOs) at 
frequencies lower than the QPO frequencies of 
normal black hole binaries by a couple of 
orders (Strohmayer \& Mushotzky 2003; Strohmayer et al. 2007). 
A straightforward interpretation of the longer timescales 
is a heavier black hole mass than $10M_\odot$. 
Hence, intermediate-mass black holes (IMBHs)
with mass $\sim 10^{2-4}M_\odot$ are considered to be one of the best
 candidates of ULXs.

The most challenging problem with such IMBHs is their formation
mechanism.  There are several possible mechanisms discussed so far.  For
instance, detailed evolution models of stellar binaries show that the
generation rate of IMBHs is very small (Madhusudhan et al. 2006).  

A promising possibility is that IMBHs could be remnants of
Population III stars (e.g., Schneider et al. 2002). 
It is suggested that zero-metalicity stars form with masses
$10^{2-3}M_\odot$ (e.g., Omukai et al. 2005).  
If the initial mass of a star is sufficiently large,
it can collapse directly to an IMBH.  These IMBHs can explain the
observed number of ULXs if they are generated in galactic disks
\citep{Krolik:2004ji}.  It is also discussed that IMBHs produced in
galactic halos can account for observed ULXs if Population III star
formation is very efficient \citep{Mii:2005as}.  This possibility is
suggested by the excess of the cosmic near-infrared background radiation
\citep{Wright:2000,Cambresy:2001,Matsumoto:2005}.  Note, however,
that with the typical baryonic fraction of Pop-III stars, 
$f\sim 10^{-5}$ \citep{Madau:2001}, 
the abundance of remnants BHs, $\Omega_{\rm BH}
=f\Omega_b\sim 4\times 10^{-7}$, would be too small to account for
IMBHs as we will see below (\S 4). On the other hand, if we assume a larger
baryonic fraction, $f \gtrsim 10^{-3}$, then the Universe may have been
reionised too early \citep{Daigne:2006}.

Black holes can be produced via the collapse
of overdense region of the early universe with large initial curvature
fluctuation \citep{Carr:1975qj}.  These black holes are called
primordial black holes (PBHs).  If the power spectrum of initial
curvature perturbation has strong peak at a specific scale, collapse
takes place at the epoch when this scale enters the horizon.  Typical
mass of resultant PBHs can be determined by the horizon mass in this
epoch.  Formation of PBHs in inflationary cosmology is discussed in
\citep{Ivanov:1997,Yokoyama:1997,Kawasaki:1998,Yokoyama:1998a,Yokoyama:1998b,Kawasaki:1999,Yamaguchi:2001}.

One exotic possibility is the formation of IMBHs during Quantum Chromo Dynamics (QCD) 
phase transition
\citep{Jedamzik:1997,Jedamzik:1998,Jedamzik:1999},
if the transition is first order. Since the equation of
state of the universe is relatively soft in this epoch, collapse of
overdense region takes place easily.  However, the horizon mass at this
epoch is only ${\mathcal{O}}(1)M_\odot$ rather than $10^{2-4}M_\odot$.

In this paper, we present a model of inflation 
\citep{Yamaguchi:2004tn,Kawasaki:2006zv}
which can produce significant amount of PBHs in astrophysically
interesting mass scales, to show that the origin of IMBHs could be such
PBHs.  In this scenario, the spectrum of initial curvature perturbation
has strong peak due to the parametric resonance
\citep{Kofman:1994,Kofman:1997,Shtanov:1995}.  After inflation, the
oscillation of the inflaton condensate can result in oscillating
effective mass of another scalar field and/or the inflaton itself.  This
oscillating effective mass excites large fluctuations of that scalar
field for specific modes.  If this scalar field contributes to the
energy density of the universe, excited fluctuation results in large
curvature perturbation for this specific modes.
If this curvature perturbation is sufficiently large,
PBHs are produced via gravitational collapse \citep{Green:2000he,Bassett:2000ha}.
In general, characteristic scales of fluctuations excited by parametric resonance
corresponds to very short wavelength compared to present observable
scale.  However, because another inflation is required after the
parametric resonance in our model, these large fluctuations are
naturally expanded to cosmologically and/or astronomically relevant
scales.

Interestingly, this scenario is related to the running spectral index of
the universe. Although it is not compelling, the result of the Wilkinson
Microwave Anisotropy Probe (WMAP) observations \citep{Komatsu:2008hk}
indicates that the running of spectral index may be large and negative.
These large running of the spectral index cannot be explained unless two
or more successive inflations are assumed
\citep{Kawasaki:2003,Easther:2006tv}. Therefore, if a large running
spectral index is confirmed by forthcoming observations of cosmic
microwave background (CMB), strongly peaked initial curvature
perturbation from multiple inflation can be an interesting candidate of
the formation mechanism of IMBH.

This paper is organized as follows. In Section 2, we explain the
scenario of inflation and how the strongly peaked initial fluctuation is
produced. The calculation of the mass distribution of PBHs is given in
Section 3. In Section 4, we estimate the amount of IMBHs produced via
the collapse of over dense region.  In Section 5, we discuss the
relation between the running spectral index and the typical mass of
PBHs.  The possibility of explaining other astronomically interesting
PBHs is also discussed.  Finally, we summarise the results.

\section{Generation of peaked power spectrum of initial curvature perturbation
via a double inflation model}

\subsection{Smooth hybrid new inflation model}

The power spectrum of initial curvature perturbation is needed to be
strongly peaked in order to result in the formation of relevant amount
of PBHs via collapse of overdense region. Initial curvature perturbation
is determined by the detail of inflation. Here we consider the smooth
hybrid new inflation model in supergravity proposed by
\citet{Yamaguchi:2004tn}. It was first considered to account for the
running spectral index suggested by WMAP.  The spectral index $n$ of the
initial curvature perturbation $\mathcal{P_R}$ is suggested to be
dependent on the comoving wavenumber $k$, according to the result of
WMAP observation.  The best-fit values of the amplitude
${\mathcal{P_R}}$, spectral index $n$ and its running $\alpha \equiv
dn/d\ln k$ at the pivot scale $k_0 = 0.002{\mathrm{Mpc}}^{-1}$ are
\begin{eqnarray}
	\label{eq:WMAP-bestfit}
	{\mathcal{P_R}}&=&(2.40\pm0.11)\times10^{-9},
 	~~n = 1.031^{+0.054}_{-0.055}, \\
	\alpha &=& -0.037 \pm 0.028 \nonumber
\end{eqnarray}
by the WMAP 5 year result only \citep{Komatsu:2008hk}.
This suggests the existence of large running from $n>1$ at long
wavelength scale to $n<1$ at short wavelength scale within the range
WMAP observation can probe, namely, $1{\mathrm{Mpc}} \lesssim k^{-1}
\lesssim 10^4{\mathrm{Mpc}}$.

In our model, two inflatons are introduced and two stages of inflation
takes place in succession. (See \citet{Kawasaki:2006zv} for the more
detailed discussions based on supergravity.) At first, the smooth hybrid
inflation occurs.  This inflation is characterized by the following
effective potential of an inflaton real scalar field $\sigma$,
\begin{eqnarray}
	\label{eq:potential-SHI}
	V(\sigma) = 
	\left\{
	\begin{array}{ll}
	 \displaystyle \mu^4 \left[ 1 - \frac{2}{27}\frac{(\mu M)^2}{\sigma^4} 
			      + \frac{\sigma^4}{8M_G^4} \right]
	 & {\mathrm{for~}}\sigma \gg (\mu M)^\frac{1}{2} \\
	 ~&~\\
	 \displaystyle \frac{8\mu^3}{M}\sigma^2 
	  & {\mathrm{for~}}\sigma \ll (\mu M)^\frac{1}{2}
	\end{array}
	\right. .
\end{eqnarray}
Here parameters with dimension of mass $\mu$ and $M$ indicate the energy
scale of smooth hybrid inflation and the cut-off scale of the underlying
theory, respectively.  Here $M_G$ is the reduced Planck mass,
$M_G=2.4\times10^{18}{\mathrm{GeV}}$. Typically, parameters $\mu$ and
$M$ satisfy $\mu \ll M \lesssim M_G$. The inflaton $\sigma$ slowly rolls
from $\sigma \gg (\mu M)^{1/2}$ to $\sigma=0$. Here you should notice
that the observational quantities ${\mathcal{P_R}}$, $n$, and $\alpha$
at the pivot scale $k_0$ of primordial fluctuations are completely
determined by the model parameters $\mu$, $M$, and $\sigma=\sigma_0$ at
the moment the scale $k_0$ exits the horizon. 
The duration of the smooth
hybrid inflation after the scale $k_0$ exits the horizon is also
determined once $\sigma_{\mathrm{ini}}$, $\mu$, and $M$ are specified.

Smooth hybrid inflation model is a variant of hybrid inflation model.
In both models inflation is driven by a false vacuum energy of
some symmetry-breaking field.  In hybrid inflation, the restored
symmetry is broken only at the end of inflation resulting in formation
of topological defects which could be cosmologically harmful.  On
the other hand, during smooth hybrid inflation, by virtue of
non-renormalizable terms, symmetry remains broken and there is no
defects formation at the end of inflation.  These non-renormalizable
terms can also serve to produce appreciable negative running spectral
index.  Furthermore, while it has been shown that hybrid inflation
requires severe fine-tuning of initial condition, \citep{Tetradis:1997wb,Mendes:2000vc}
smooth hybrid inflation is free from such a fine tuning and can be realized naturally.

The most important feature of this inflation model is
that the largelly running spectral index can be realized.
This potential has positive curvature for larger $\sigma$ and negative
curvature for smaller $\sigma$. Since the spectral index of the initial
curvature perturbation depends on the gradient of the potential, this
inflation results in scale-dependent spectral index. Large wavelength
modes of the initial curvature perturbation are produced while the
inflaton takes a large value. Therefore, the spectral index for these
modes is larger than unity. On the other hand, short wavelength modes
are produced while the inflaton takes a smaller value. Thus, the
spectral index for these modes is smaller than unity. Consequently,
negative running of the spectral index suggested by the WMAP 5 year
result can be realized. 
For example, a parameter choice
(equations are corrected)
\begin{eqnarray}
	\mu = 2.1\times10^{-3}M_G,~M=1.3M_G,~\sigma_0=0.227M_G
\end{eqnarray}
gives
\begin{eqnarray}
	\label{eq:WMAP-good}
	{\mathcal{P_R}}&=&2.40\times10^{-9},
 	~~n = 1.040, \\
	\alpha &=& -0.033, \nonumber
\end{eqnarray}
which is in the range (\ref{eq:WMAP-bestfit}).
However, since large running needs large
variation of the gradient of the potential, the evolution of inflaton
must be very fast in order to account for the large running of the
spectral index, which implies that smooth hybrid inflation does not last
so long. Thus, another inflation is necessary to push the relevant
scales to cosmologically observable scale.

In this model, new inflation is considered as another inflation to
compensate for the duration of the inflationary epoch. Moreover, it
naturally leads to low reheating temperature to avoid the overproduction
of the gravitinos. The effective potential of the second inflaton field
$\phi$ during this new inflation is given by
\begin{eqnarray}
	V(\phi) = v^4 - \frac{c}{2}\frac{v^4\phi^2}{M_G^2} 
	 - \frac{g}{2}\frac{v^2\phi^4}{M_G^2}
	+ \frac{g^2}{16}\frac{\phi^8}{M_G^4},
\end{eqnarray}
where $v$ determines the scale of new inflation, which is assumed to be
$v \ll \mu$.  Parameters $c \lesssim 1$ and $g < 1$ determine the shape
of the potential $V(\phi)$.  In addition, the second inflaton field
$\phi$ is coupled to $\sigma$ through gravitationally suppressed
interactions in supergravity even if no direct coupling is introduced.
Since the contribution from this coupling dominates the potential of
$\phi$, $\phi$ is trapped to a value close to, but different from
 $\phi=0$, while $\sigma$ has a non-vanishing
value.  After the smooth hybrid inflation, $\sigma$ oscillates around
$\sigma=0$ with its amplitude decreasing.  Eventually, the contribution
from $V(\phi)$ becomes dominant, and then $\phi$ begins to roll slowly
from the vicinity of $\phi=0$ to the minimum $\phi \simeq
(4v^2M_G^2/g)^{1/4}$. The duration of new inflation is determined by
parameters $v$, $g$, and $c$. Note that fluctuation generated during the
new inflation is not constrained by observations unless it is extremely
large, because length scale of such fluctuation is too small to be
cosmologically observable. The exception is the strong fluctuation
generated via the parametric resonance \citep{Kawasaki:2006zv}, which
may serve as a seed of PBHs, as discussed below.

\subsection{Generation of strong fluctuation via the parametric resonance}

Because of the self-coupling of $\sigma$, oscillating condensate of
$\sigma$ gives oscillating contribution to its own effective mass.
Since this self-coupling is large, the amplitude of this oscillating
contribution is larger than its mass at the origin $m_\sigma \simeq
(8\mu^3/M)^{1/2}$.  Hence some specific modes of fluctuation of $\sigma$
are strongly amplified via the parametric resonance.  In our case, it is
determined by the evolution equation of Fourier modes of fluctuation
$\sigma_k$, which can be approximated as
\begin{eqnarray}
	\label{eq:mathieu}
	\sigma_k'' + \left[ A - 2q\cos(2z) \right]\sigma_k =0.
\end{eqnarray}
Here $A$ is a function of $k/(a m_\sigma)$, where $a$ is the scale
factor.  $q$ is a time-dependent parameter proportional to the
self-coupling and the amplitude of the oscillation of the zero-mode of
$\sigma$.  The prime represents derivative with respect to the time
variable $z$ defined by the proper time as $2z = m_\sigma t - \pi/2$.
Equation (\ref{eq:mathieu}) has the shape of Mathieu equation.  The
solution of this equation have infinite sequence of instability bands
determined by the value of $A$.  In these instability bands, the
solution grows exponentially for $q\gtrsim 1$.  Hence, the modes $k$
satisfying the condition $k/a = m_\sigma \times (const.)$ grows
exponentially.  According to numerical calculation, the strongest
instability appears at around $k/a =
m_\sigma\times{\mathcal{O}}(10^{-1})$.  Note that the comoving scale $k$
in this instability band is inside the horizon at the beginning of the
new inflation.  Eventually $q$ decreases to $q\ll1$ and the parametric
resonance ceases.  Later, fluctuations $\phi_k$ with the same range of
$k$ get large amplitudes via linear coupling with the amplified
fluctuation $\sigma_k$.  These fluctuations $\phi_k$ determine the
curvature perturbation during the new inflation.  Thus, the
curvature perturbation has a large amplitude at a 
specific range of $k$.

We can see that the position of the strong peak $k_p$ is approximately determined by
the factor $(\mu M)$ and the duration of the smooth hybrid inflation.
The peak position $k_p$ satisfies
\begin{eqnarray}
	\label{eq:peak_position}
	\frac{k_p}{a_e} \simeq C m_\sigma,
\end{eqnarray}
where $C \simeq \mathcal{O}(10^{-1})$ is determined by the 
detail of the parametric resonance and $a_e$ is the
scale factor at the end of the smooth hybrid inflation.  On the other
hand, the scale factor $a_0$ at the moment the pivot scale exits the
horizon and the Hubble parameter during the smooth hybrid inflation, $H_s
\sim \mu^2/(\sqrt{3}M_G)$, satisfies
\begin{eqnarray}
	\label{eq:pivot_position}
	\frac{k_0}{a_0 H_s} =1.
\end{eqnarray}
Therefore, we get
\begin{eqnarray}
	\label{eq:pivot_peak}
	\ln \frac{k_p}{k_0} = \ln \frac{a_e}{a_0} + \ln\frac{ M_G}{\sqrt{\mu M}} + \ln C + 1.6.
\end{eqnarray}
The shorter smooth hybrid inflation
we assume, the larger-mass PBHs are produced.

Note that the typical mass of PBHs is determined once we specified 
${\mathcal{P_R}}$, $n$ and $\alpha$.
In our model, only the smooth-hybrid inflation is responsible to the curvature perturbation
of cosmologically interesting scales.
Since free parameters of our smooth-hybrid inflation model are
$\mu$, $M$ and the value of inflaton $\sigma_0$ at the moment the pivot scale $k_0$
crosses outside the horizon,
cosmological parameters ${\mathcal{P_R}}$, $n$ and $\alpha$
depend only on parameters $\mu$, $M$ and $\sigma_0$.
Therefore, in order to reproduce specific values of ${\mathcal{P_R}}$, $n$ and $\alpha$,
we look for appropriate values of $\mu$, $M$ and $\sigma_0$.
Once we find them,
$a_e/a_0$ is determined by the dynamics of the inflaton,
and then $k_p/k_e$ is specified by the Eq.(\ref{eq:pivot_peak}).

The amount of PBHs depends on the height of the peak, which is
determined by the balance of the efficiencies of the parametric
resonance and the subsequent decay of $\sigma$ into lighter particles.
If this decay rate is sufficiently large, it makes the parametric
resonance weaker. It is extremely difficult to estimate the magnitude of
this strong peak analytically because of the non-perturbative nature of
the parametric resonance. 

\subsection{Spectrum of the curvature perturbation for the formation of IMBHs}

\begin{figure}
	\begin{center}			
		\includegraphics[width=.9\linewidth]{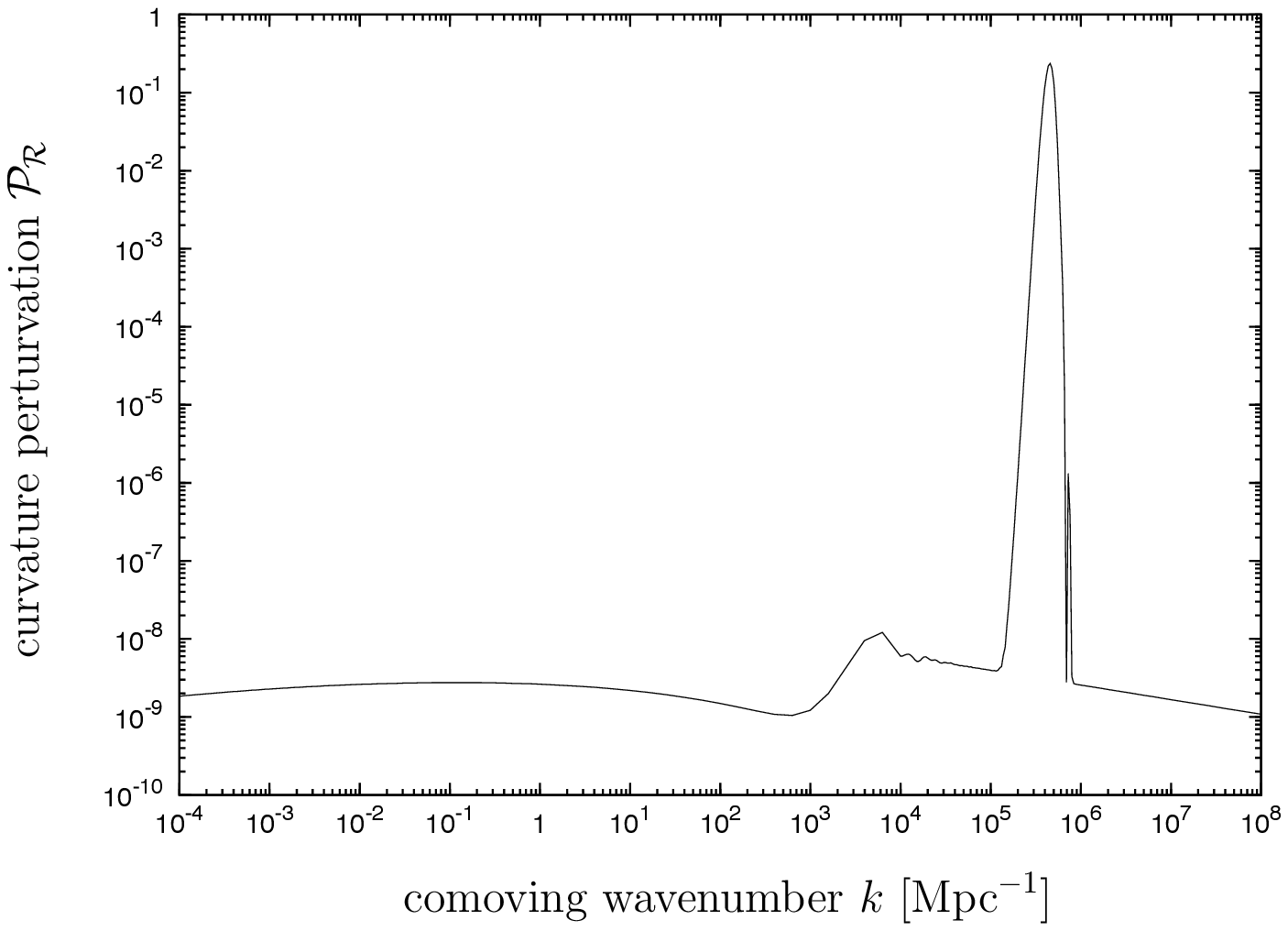}
		\caption{\small 
		The spectrum leading to the formation of IMBHs
		from the smooth hybrid new inflation model.
		}
	\label{fig:spectrum}
	\end{center}
\end{figure}

In Fig.\ref{fig:spectrum}, we show an example of the spectrum of the
initial curvature perturbation which results in the formation of
significant amount of IMBHs. 
In order to give a example of generation of $10^2 M_\odot$ PBHs,
we choose $\mu=1.6\times10^{-3}
M_G,~M=0.81M_G,~v=\mu/4,~c=0.1$, and
$g=10^{-6}$ for parameters of the smooth hybrid new inflation as an
example. The position of the peak is at $k_p = 4.4\times10^5
{\mathrm{Mpc}}^{-1}$, thus the typical mass of resultant black holes is
estimated as $M_{\mathrm{PBH}} = 108 M_\odot$ by the horizon mass at
the moment the scale $k_p$ enters the horizon.  
Unfortunately, it is difficult to estimate the range of these parameters
which results $M_{\mathrm{PBH}} \sim 100 M_\odot$.
As a hint for this estimation, we tried various parameters for $\mu$ and $M$.
For example, $\mu=1.7\times10^{-3}M_G$ and $M=0.86M_G$ gives 
$M_{\mathrm{PBH}} \simeq 1000 M_\odot$,
while $\mu=1.5\times10^{-3}M_G$ and $M=0.72M_G$ gives 
$M_{\mathrm{PBH}} \simeq 10 M_\odot$,
both reproducing the cosmological parameters consistent with (\ref{eq:WMAP-bestfit}).
We also assumed the
decay rate of $\sigma$ as $\Gamma = 7.65\times10^{-8} M_G$.  Modes of
curvature perturbation with $k<10^3{\mathrm{Mpc}}^{-1}$ are generated
during the smooth hybrid inflation. This spectrum gives
${\mathcal{P_R}}=2.4\times10^{-9},~n=1.067$, and $\alpha=-0.014$ at
$k_0=0.002{\mathrm{Mpc}}^{-1}$. 
A large and negative running spectral index is a
characteristic feature of formation of IMBHs via this scenario, which
can be tested by forthcoming observations of CMB.

\section{Formation of primordial black holes via collapse of overdense region}

PBHs were considered to be produced via gravitational collapse if the
density perturbation $\delta \equiv \delta\rho/\rho$ within a patch of
the universe is larger than some critical value $\delta_c$ at the moment
the Hubble radius becomes smaller than the size of the patch
\citep{Carr:1975qj}.  Later, a critical phenomenon was observed
\citep{Niemeyer:1997mt} in the formation of PBHs via gravitational
collapse of radiation fluid in the Friedmann-Robertson-Walker background
\citep{Evans:1994pj}.  According to \citep{Niemeyer:1997mt}, the mass
of PBH $M_{\mathrm{PBH}}$ produced via collapse of a overdense region
with the density perturbation $\delta$ has a scaling relation
\begin{eqnarray}
	\label{eq:scaling-_relation}
	M_{\mathrm{PBH}} = \kappa M_H (\delta - \delta_c)^\gamma,
\end{eqnarray}
where $M_H$ is the horizon mass at the moment this region enters the horizon.
The index $\gamma$ and the critical value $\delta_c$ are universal constants,
which are estimated by numerical simulations as $\gamma\simeq0.35$ and 
$\delta_c \simeq 0.67$, where the latter is quoted from results of
numerical simulation of critical collapse\citep{Niemeyer:1999ak}.

Given strongly peaked spectrum of curvature perturbation, 
the mass spectrum of PBHs is estimated as follows \citep{Yokoyama:1998xd,Barrau:2002ru}.
Assuming that the initial curvature perturbation has a Gaussian
probability distribution and that the curvature perturbation is strongly
peaked, the differential mass spectrum of the PBHs at the time of
formation can be given by.
\begin{eqnarray}
	\label{eq:PBH_spectrum}
	\frac{d \Omega_{\mathrm{PBH}}(t_c)}{d \ln M_{\mathrm{PBH}}}
	= \epsilon^{-\frac{1}{\gamma}}\beta(M_H) \left( 1 + \frac{1}{\gamma} \right)
	\left( \frac{M_{\mathrm{PBH}}}{M_H} \right)^{1 + \frac{1}{\gamma}}~~~~~~~~\nonumber \\
	\times\exp\left[ -\epsilon^{-\frac{1}{\gamma}} (1+\gamma) 
	\left( \frac{M_{\mathrm{PBH}}}{M_H} \right)^\frac{1}{\gamma} \right].
\end{eqnarray}
The parameter $\epsilon$ is defined as the ratio of the peak of
differential mass spectrum $M_{\mathrm{max}}$ to the horizon mass $M_H$.
The function $\beta(M_H)$ is the probability that the relevant mass
scale has an above-threshold amplitude of fluctuations to collapse as it
enters the Hubble radius, namely,
\begin{eqnarray}
	\label{eq:beta}
	\beta(M_H) = \int_{\delta_c} p(\delta,t_c)d\delta 
	\simeq \frac{\sigma_H(t_c)}{\sqrt{2\pi}\delta_c} \exp\left( -\frac{\delta_c^2}{2\sigma_H^2(t_c)} \right) 
\end{eqnarray}
where $p(\delta,t_c)$ is the probability distribution function of the
density perturbation at the moment the scale $k_p$ enters the particle
horizon, which is assumed to be Gaussian.  The variance of the density
perturbation $\sigma_H(t_c)$ smoothed over the comoving length scale
$k_p^{-1}$ can be estimated by
\begin{eqnarray}
	\label{eq:variance}
	\sigma_H^2(t_c) = \frac{16}{81} \int_p \left(\frac{k}{k_p}\right)^3 {\mathcal{P_R}}(k) 
	T^2\left( \frac{k}{k_p},t_c \right) W_{\mathrm{TH}}^2 \left(\frac{k}{k_p}\right) \frac{dk}{k_p}.
\end{eqnarray}
Here we employed the top-hat window function
\begin{eqnarray}
	\label{eq:top-hat}
	W_{\mathrm{TH}}(x) = \frac{3}{x^2}\left( \frac{\sin x}{x} - \cos x \right).
\end{eqnarray}
Note that the power spectrum of the density perturbation ${\mathcal{P}}_\delta$ is given by
\begin{eqnarray}
	\label{eq:curv-dens}
	{\mathcal{P}}_\delta = \frac{16}{81} \left( \frac{k}{k_p} \right)^4 T^2\left( \frac{k}{k_p},t_c \right)
	{\mathcal{P_R}},
\end{eqnarray}
where $T(k/k_p,t_c)$ is the transfer function which is given by
\begin{eqnarray}
	\label{eq:transfer}
	T \left( \frac{k}{k_p},t_c \right) = 
	\frac{k_p^2}{k^2}\left\{ \frac{\sqrt{3}k_p}{k}\sin\left( \frac{k}{\sqrt{3}k_p} \right) 
	 - \cos\left( \frac{k}{\sqrt{3}k_p} \right) \right\}.
\end{eqnarray}

\begin{figure}
	\begin{center}			
		\includegraphics[width=.9\linewidth]{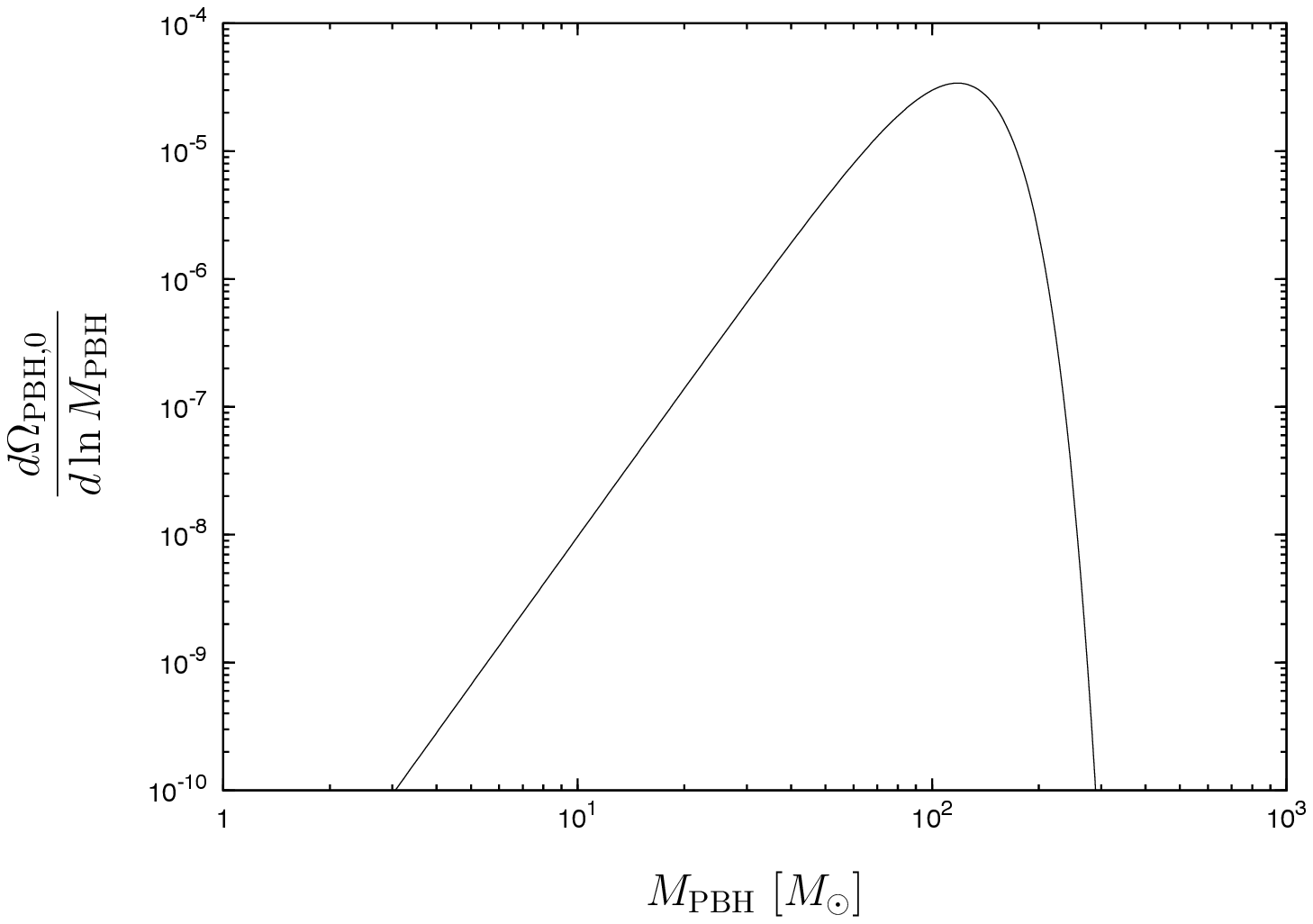}
		\caption{\small 
		The differential mass spectrum (\ref{eq:massspec-pre}) calculated for the spectrum
		shown in Fig.\ref{fig:spectrum}.
		}
	\label{fig:mass_spectrum}
	\end{center}
\end{figure}

Finally, the differential mass spectrum of PBHs at present can be given by 
\begin{eqnarray}
	\label{eq:massspec-pre}
	\frac{d \Omega_{\mathrm{PBH},0}}{d \ln M_{\mathrm{PBH}}} = 
\frac{d \Omega_{\mathrm{PBH}}(t_c)}{d \ln M_{\mathrm{PBH}}}
\exp\left( \int_{t_c}^{t_0}3wHdt\right)
\end{eqnarray}
unless PBHs are dominant constituent of dark matter, 
where $w$ is the effective equation-of-state parameter of the total
cosmic energy density including contribution from dark energy.
In Fig.\ref{fig:mass_spectrum}, we show the
differential mass spectrum (\ref{eq:massspec-pre}) calculated for the
spectrum shown in Fig.\ref{fig:spectrum} with $\epsilon=1$.  
We took contribution from dark energy into account.
We can see
that the mass distribution of PBHs is well peaked around
$M_{\mathrm{PBH}}=108M_\odot$.  Therefore, we can conclude that only
IMBHs are produced from this spectrum.

\section{Amount of IMBHs}

The observational constraint on the total amount of IMBHs can be
estimated from the observed number density of ULXs $10^{-3.5}
{\mathrm{Mpc}}^{-3}$, as follows.  Assuming that ULXs are IMBHs with
typical mass $10^2M_\odot$, this number density means that the
contribution of IMBHs observed as ULXs to the density parameter of the
universe $\Omega_{\mathrm{ULXBH}}$ is $\sim10^{-12.5}$.  
Agol and Kamionkowski (2001) estimate  the probability 
for floating black holes to acquire high enough accretion rates 
from interstellar medium as $\sim10^{-5}$.
Furthermore, if we consider the enhancement of 
dark matter (and IMBHs) in the galactic disk region properly,
the typical mass fraction 
of dark matter (over the entire dark matter halo) within the 
galactic disk region would be $\sim10^{-3}$ 
(see Mii \& Totani 2005). 
By combining the above three points, 
we can then estimate the total abundance of the IMBHs in the
universe as
\begin{eqnarray}
 \label{eq:abundance}
	\Omega_{\mathrm{IMBH}} \sim \Omega_{\mathrm{ULXBH}}
 	\times (10^{-5}\times10^{-3})^{-1}
	= 10^{-4.5}.
\end{eqnarray}
We note that estimations adopted above would be valid only up to 
the order of magnitude.

We here point out that $\Omega_{\mathrm{IMBH}}$ quoted satisfies 
the constraints from CMB observations 
(see Ricotti, Ostriker \& Mack 2007). 
Namely, if there are too many PBHs, a fraction of these black holes 
shining by gas accretion in the early universe at $z\approx 100-1000$ 
(where gas density is quite high) must have distorted the CMB spectrum.
Then, the CMB spectral distortions (based on Far-Infrared Absolute Spectrophotometer (FIRAS) data) provide 
an upper limit for allowed $\Omega_{\mathrm{IMBH}}$ to be 
$10^{-4.5}$ for the case of the PBH mass $M_{\mathrm{PBH}}=10^2 M_\odot$ 
(Ricotti et al. 2007). 

\begin{figure}
	\begin{center}			
		\includegraphics[width=.9\linewidth]{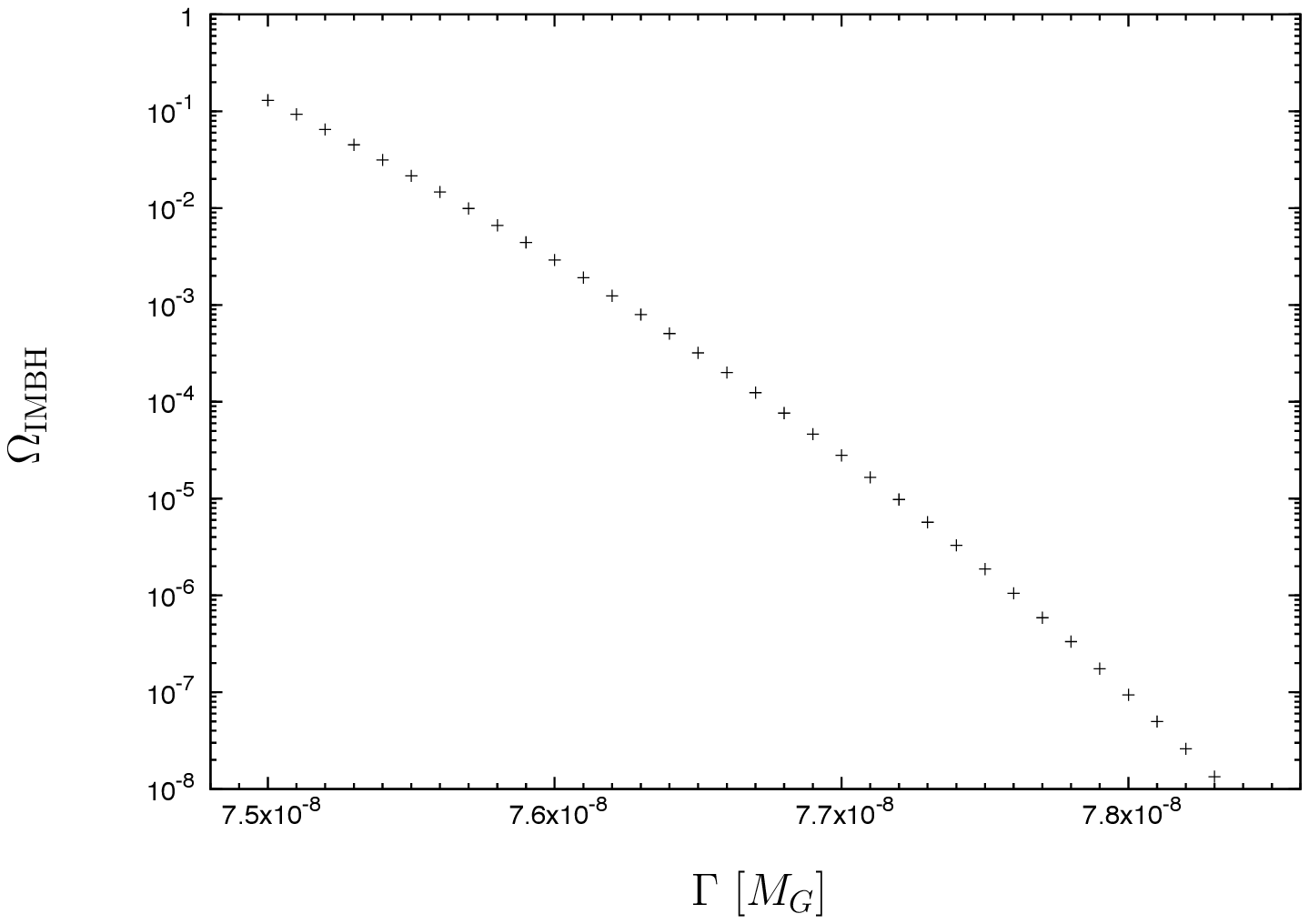}
		\caption{\small 
		The relation between the abundance of IMBHs $\Omega_{\mathrm{IMBH}}$ and
		the decay rate $\Gamma$.
		}
	\label{fig:Gamma-Omega}
	\end{center}
\end{figure}

We calculated the abundance of the IMBHs from the differential mass
spectrum Eq.(\ref{eq:massspec-pre}) for the same parameter set as we
chose for Fig.\ref{fig:spectrum} and various decay rates $\Gamma$.  The
result is summarised in Fig.\ref{fig:Gamma-Omega}, which shows that the
abundance of IMBHs is sensitively dependent on the decay rate $\Gamma$.
Accordingly, the range $7.65\times10^{-8}M_G < \Gamma <
7.74\times10^{-8}M_G$ results in the abundance $10^{-5.5} <
\Omega_{\mathrm{IMBH}} < 10^{-3.5}$.  Thus, this scenario needs tuning
of the decay rate in order to account for the abundance of ULXs.  
This is simply due to the fact that the abundance of PBHs is
exponentially sensitive to the peak amplitude of fluctuation, and all
the other models of PBH formation requires the same level of fine tuning.
Note that the decay rate of $\sigma$ is expected to be given by $\Gamma
\simeq N g_\sigma^2 m_\sigma/(8\pi)$ where $g_\sigma$ is some gauge
coupling constant included in near-Planck scale physics and $N$ is the
number of decay channels. Under the parameter set we have chosen, the
above constraint on $\Gamma$ is rewritten by $Ng_\sigma^2 \sim 10^{-2}$, which
can be satisfied by natural values of $N$ and $g_\sigma$. Here one
should notice that $\Gamma$ has nothing to do with the reheating
temperature after the final
inflation, which is determined by the decay rate
of $\phi$.

\section{Spectral index and typical mass of PBHs}

Here let us consider the relation between the typical mass of PBHs and
cosmological parameters, spectral index and its running.  As we
discussed in the Section 2, ${\mathcal{P_R}}$, $n$ and $\alpha$ at the
pivot scale $k_0 = 0.002{\mathrm{Mpc}}^{-1}$ are determined once we
choose parameters $\mu,~M$ and the value of $\sigma$ at the moment the
scale $k_0$ exits the horizon.  This choice simultaneously determines
the duration of the smooth hybrid inflation $\ln (a_e/a_0)$.  Thus, if
we fix the amplitude ${\mathcal{P_R}}$ to be the best-fit value of the
result of WMAP3, we can compute the mass of PBH for each combination of
$(n,\alpha)$, up to an uncertainty in the detail of parametric resonance.
A brief explanation of the method we used is given in Appendix.

\begin{figure}
	\begin{center}			
		\includegraphics[width=.9\linewidth]{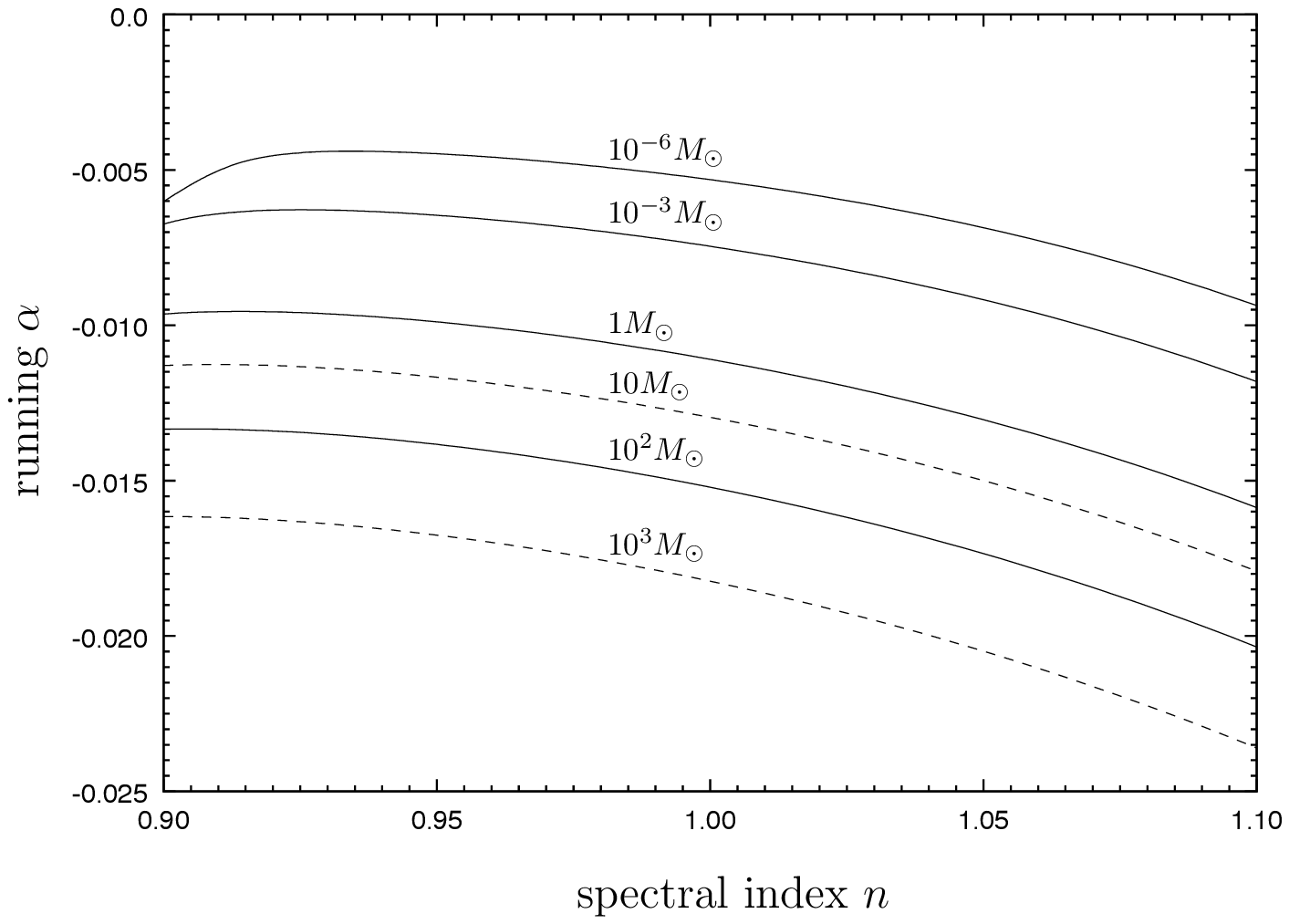}
		\caption{\small 
		Contour-lines of the typical mass of PBHs 
		on the $n$-$\alpha$ plane.
		}
	\label{fig:contour}
	\end{center}
\end{figure}

In Fig.\ref{fig:contour}, we show the relation\footnote{
Cosmological parameters $n$ and $\alpha$ of the spectrum shown in
Fig.\ref{fig:spectrum} and the typical mass of PBHs $108M_\odot$ is
deviated from this result.  The main reason is that $v$ is not
negligibly small compared with $\mu$.  Since long oscillatory phase
requires extremely long numerical calculation, we chose large $v$ for
our calculation.
} between typical mass of PBHs and the cosmological parameters $n$ and
$\alpha$.  We can see that larger mass PBHs requires larger running of
the spectral index, while lower mass PBHs can be produced with negligible
running.  This can be understood by the fact that the short smooth
hybrid inflation is realized for steep potential, which means large
running.  
If we choose some other appropriate values of model parameters,
this scenario can alternatively explain one of other astronomically interesting PBHs,
such as $0.1-1M_\odot$ BHs considered to be a candidate of
massive compact halo objects (MACHOs) \citep{Alcock:2000ph} 
or $10^{15}{\mathrm{g}}$ BHs which
are possible sources of ultra high energy cosmic rays (UHECRs) \citep{Takeda:2002at},
although the latter possibility is probably no longer interesting given the recent Auger data
\citep{Aglietta:2007yx}. 
The former requires $\alpha\sim -0.01$, while the latter requires negligibly
small $\alpha$.  For the latter case, best-fit values of
${\mathcal{P_R}}$ and $n$  is given by
\begin{eqnarray}
	\label{eq:WMAP-bestfit2}
	{\mathcal{P_R}}=(2.41\pm0.11)\times10^{-9},~~n = 0.963^{+0.014}_{-0.015}
\end{eqnarray}
from the result of WMAP 5 year only,
assuming $\alpha=0$, which can also be realised in the present scenario.

\section{Summary and discussion}

We considered the formation of IMBHs as a candidate of ULXs via
gravitational collapse of overdense region of the curvature perturbation
whose spectrum is strongly peaked.  This initial fluctuation is produced
by a double inflation model, the smooth hybrid new inflation.  The peak
is originated from the amplification of specific modes of fluctuation
via the parametric resonance.  This inflation model can also explain the
observed running spectral index.  We found the relation between typical
mass of PBHs and the cosmological parameters $n$ and $\alpha$.  This
result shows that the formation of IMBHs in this scenario requires
significant running of the spectral index.  The amount of IMBHs depends
on the height of the peak, which is determined by the detail of
parametric resonance and the decay rate of the inflaton.  The decay rate
must be some specific and fine-tuned 
value for required abundance of IMBHs. 
The level of required fine-tuning is no different from other models
of PBH formation.  What is important here is that the 
 requirement can be satisfied with natural values of the 
parameters. 
This scenario can alternatively explain one of other PBHs, 
such as $0.1-1M_\odot$ BHs considered to be
candidates of MACHOs or $10^{15}{\mathrm{g}}$ BHs which are possible
sources of UHECRs
with some other appropriate choices of model parameters.

The production of strongly peaked initial curvature perturbation via the
parametric resonance can take place in other double inflation models
with intermediate oscillatory phases
\citep{Kawasaki:2003,Yamaguchi:2003}. Since these double inflation
models are required to explain the possible large running of the
spectral index, observational confirmation of large and negative running
spectral index may be regarded as a hint of IMBHs.

\section*{Appendix}
Here we will explain how Fig.\ref{fig:contour} was determined.
We used analytic estimations given in \citep{Yamaguchi:2004tn},
because it is sufficient to see approximate dependence of $M_{\mathrm{PBH}}$ on $(n,\alpha)$.

According to \citep{Yamaguchi:2004tn}, the dynamics of inflaton $\sigma$
has characteristic values of $\sigma$.
Namely, $\sigma_d$, where the dominant contribution to the dynamics
changes from the second term in Eq.(\ref{eq:potential-SHI}) to the third term,
and $\sigma_c$, where the slow-roll condition is violated and the smooth hybrid inflation ends.
These values are estimated by following equations.
Note that we consider only the case $m=2$ in \citep{Yamaguchi:2004tn}.
\begin{eqnarray}
	\label{eq:sigma_d}
	\sigma_d &=& \left(\frac{16}{27}\right)^\frac{1}{8}\times\left(\frac{\mu M}{M_G^2}\right)^\frac{1}{4}, \\
	\label{eq:sigma_c}
	\sigma_c &=& \left(\frac{40}{27}\right)^\frac{1}{6}\times\left(\frac{\mu M}{M_G^2}\right)^\frac{1}{3}.
\end{eqnarray}
The number of $e$-foldings $N_d$ from $\sigma=\sigma_d$ to $\sigma=\sigma_c$ is
\begin{eqnarray}
	\label{eq:N_d}
	N_d = -\frac{5}{6}+\left(\frac{1}{48}\right)^\frac{1}{4}\times\left(\frac{\mu M}{M_G^2}\right)^\frac{1}{2}.
\end{eqnarray}

Now, if we choose the value of $\mu M$ and 
require the typical mass of resultant PBHs $M_{\mathrm{PBH}}$,
$N_e\equiv \ln (a_e/a_0)$ is determined by Eq.(\ref{eq:pivot_peak}).
Then, the value of $\sigma_0$, which is the value of $\sigma$ at the moment the pivot scale
$k_0$ crosses outside the horizon, is determined by tracing back the dynamics of $\sigma$
from the end of smooth hybrid inflation $\sigma=\sigma_c$.
If $N_e>N_f$, it means that $\sigma_0 > \sigma_d$. Therefore,
\begin{eqnarray}
	\label{eq:sigma0-1}
	\sigma_0 = \left[ \left\{ \left( \frac{1}{48} \right)^\frac{1}{4} 
	+ \left( \frac{27}{16} \right)^\frac{1}{4}  \right\}\left(\frac{\mu M}{M_G^2}\right)^{-\frac{1}{2}} - 
	\frac{5}{6} - N_e \right]^{-\frac{1}{2}}.
\end{eqnarray}
On the other hand, if $N_e<N_f$, then $\sigma_0 < \sigma_d$ and
\begin{eqnarray}
	\label{eq:sigma0-2}
	\sigma_0 = \left( \frac{5}{6} + N_e \right)^\frac{1}{6}\left(\frac{4}{3}\right)^\frac{1}{3}
	\left(\frac{\mu M}{M_G^2}\right)^\frac{1}{3}.
\end{eqnarray}
Thus, we can determine $\sigma_0$ if we specify $(\mu M)$ and $M_{\mathrm{PBH}}$.

Because ${\mathcal{P_R}}$ is given by
\begin{eqnarray}
	\label{eq:P_R-analy}
	{\mathcal{P_R}} = \frac{\mu^2 M_G}{\sqrt{3}\pi}\left\{ \sigma_d^3\left( \frac{\sigma_d}{\sigma_0} \right)^5
	 + \sigma_0^3 \right\}^{-1},
\end{eqnarray}
parameters $\mu$, and then $M$, can be calculated once we determined $\sigma_0$.
Cosmological parameters $n$ and $\alpha$ are calculated via slow-roll parameters.

Consequently, if we choose $M_{\mathrm{PBH}}$,
we can draw a curve on $(\mu,M)$ plane which results the typical mass of PBHs $M_{\mathrm{PBH}}$.
Deriving the corresponding values of $(n,\alpha)|_{\sigma=\sigma_0}$,
we can draw Fig.\ref{fig:contour}.

\section*{Acknowledgments}

JY is grateful to Yudai Suwa for useful communications.  This work was
partially supported by JSPS Grant-in-Aid for Scientific Research
No.~19740105(TK), No.~18540254(MK), No.~18740157(MY), No.~16340076, and
No.~19340054(JY), and JSPS research fellowships (TT). 
This work was also supported in part by JSPS-AF
Japan-Finland Bilateral Core Program (MK) and World Premier International
Research Center InitiativeiWPI Initiative), MEXT, Japan (MK and JY).

\label{lastpage}

\end{document}